\documentclass[twocolumn,groupedaddress]{revtex4-1}
\usepackage{dcolumn}
\usepackage{graphicx}
\usepackage{color}

\begin{document}


\title{Truncated correlation hierarchy schemes \\  for driven-dissipative multimode quantum systems}


\author{W. Casteels$^1$, S. Finazzi$^{1}$, A. Le Boit\'e$^{1,2}$, F. Storme$^1$, C. Ciuti$^1$}
\affiliation{$^1$Laboratoire Mat\'eriaux et Ph\'enom\`enes Quantiques, Universit\'{e} Paris Diderot, Sorbonne Paris Cit\'{e} and CNRS, UMR 7162,  75205 Paris Cedex 13, France}	
\affiliation{$^2$Institut f\"ur Theoretische Physik and IQST, Albert-Einstein-Allee 11, Universit\"at Ulm, 89069 Ulm, Germany}


\date{\today}

\begin{abstract}
We present a method to describe driven-dissipative multi-mode systems by considering a truncated hierarchy of equations for
the correlation functions.  We consider two hierarchy truncation schemes with a global cutoff on the correlation order, which is the sum of the exponents of the operators involved in the correlation functions:  a 'hard' cutoff corresponding to an expansion around the vacuum, which applies to a regime where the number of excitations per site is small; a 'soft' cutoff which corresponds to an expansion around coherent states, which can be applied for large excitation numbers per site. This  approach is applied to describe the bunching-antibunching transition in the driven-dissipative Bose-Hubbard model for photonic systems. The results have been successfully benchmarked by comparison with calculations based on the corner-space renormalization method in 1D and 2D systems. The regime of validity and strengths of the present truncation methods are critically discussed. 
\end{abstract}


\maketitle

\section{Introduction}
Recently there has been a strong interest in driven-dissipative manybody systems, particularly non-equilibrium photonic systems. These systems can be realized with various platforms, including semiconductor microcavities and superconducting circuits  (see for example Refs. \cite{RevModPhys.85.299, ANDP:ANDP201200261, RevModPhys.86.1391} for recent reviews). Typically, the steady-state of such systems can be far from equilibrium due to the competition between driving and dissipation, which  can not be neglected. In many regimes, they can be described by a Lindblad master equation for the manybody density-matrix. The possibility of coupling many of these  modes (e.g., in lattices of coupled nonlinear resonators) is currently receiving a lot of attention for the realization of quantum many-body states of light. 

Due to the inherent out-of-equilibrium nature of these systems many of the numerical approaches developed for systems at equilibrium are not necessarily applicable. These systems are typically computationally more challenging since the total number of photons is typically not conserved resulting in a much larger effective Hilbert space. Moreover, they are in general in a mixed state described by a density matrix while a closed system at low temperature is described by the groundstate wavefunction. As a result, exact diagonalisation approaches are only applicable for relatively small systems. Examples of numerical tools for larger systems are the TEBD algorithm \cite{PhysRevLett.98.070201, PhysRevB.78.155117} and a DMRG-type variational approach for the steady-state \cite{PhysRevLett.114.220601, PhysRevA.92.022116}. The underlying structure of both these techniques is based on the so-called matrix product states:  such approaches are efficient in 1D,  but encounter difficulties for 2D systems \cite{doi:10.1080/14789940801912366, Schollwock201196}. For the description of  2D lattices, the corner-space renormalization method has been recently introduced \cite{PhysRevLett.115.080604}. 

In this paper we investigate truncation schemes for the infinite hierarchy of equations coupling correlations functions. Both schemes are based on a global cutoff on the correlation order, that is the sum of the exponents of the operators involved in the correlation functions. In particular, one truncation scheme corresponds to an expansion around the vacuum (hard cutoff) and is most appropriate when the number of excitations is small. The second truncation scheme is
based on a self-consistent 'soft' cutoff, which corresponds to an expansion around coherent states and applies when the number of excitations per site is large. The numerical implementation of this scheme is straightforward with respect to the more advanced numerical approaches discussed above. Moreover, it allows to explore regimes that are very hard to reach with techniques that are based on a cutoff in the number of photons, such as exact diagonalization. The presented numerical scheme is very versatile and allows to study various theoretical models with a wide range of applications. Recently a scheme with a hard cutoff has been applied in Ref. \cite{0953-4075-46-22-224023} to obtain a perturbative expansion in the hopping for an infinite 2D driven-dissipative Bose-Hubbard lattice. Note that the truncation schemes presented in the present paper are different from the approach in Ref. \cite{0953-4075-46-22-224023} where a local cutoff for the exponents of the operators is enforced for each site separately.

As an example we apply the method to the driven-dissipative Bose-Hubbard model. In particular the physics of the anti-bunching to bunching transition is examined: such an effect was investigated for or a single resonator (e.g. in Ref. \cite{PhysRevB.73.193306}) and explored more recently in Ref. \cite{PhysRevA.93.023821} for a lattice of dissipative two-level systems. Both a 1D chain as well as a 2D square lattice of resonators are investigated and the results are compared with the results obtained from the corner-space renormalization method. Two sets of system parameters are considered, one resulting in a relatively small photon density, ideally suited for the scheme with a hard cutoff, and one with a relatively high density which is well captured by the soft cutoff truncation. In all cases, the presence of additional modes is shown to reduce both the bunching and the anti-bunching and for finite systems an oscillation of the second order correlation functions as a function of the cavity-drive frequency detuning is shown. As the size of the system is increased these oscillations become less pronounced and eventually vanish in the thermodynamic limit, resulting in an overall bunching to anti-bunching transition at zero detuning, corresponding to the middle of the single-particle energy band.

This paper is structured as follows:  in Section II we introduce the theoretical approach in a general way;  in Section III, we specialize the theory to the driven-dissipative Bose-Hubbard model;  in Section IV the numerical results are presented and critically discussed. Finally, in Section V, the conclusions are presented together with an outlook regarding the perspectives of the method.

\section{Theoretical framework and description of the method}

\subsection{The Lindblad master equation}
If a system is weakly interacting with a large bath, such that retardation effects can be neglected and the Markov approximation is valid, the degrees of freedom of the environment can be traced out to get an effective Lindblad master equation for the reduced density operator  $\hat{\rho}$ of the system (we set $\hbar = 1$):
\begin{equation}
\frac{\partial\hat{\rho}}{\partial t}=i\left[\hat{\rho},\hat{H}\right] + \sum_j\left[\hat{C}_j\hat{\rho}\hat{C}^\dagger_j - \frac{1}{2}\left(\hat{C}_j^\dagger\hat{C}_j\hat{\rho}+\hat{\rho}\hat{C}_j^\dagger\hat{C}_j\right) \right],
\label{eq:Master}
\end{equation}
where $\hat{H}$ is the system Hamiltonian and $\hat{C}_j$ are the jump operators that describe the interaction with the environment. From the master equation (\ref{eq:Master}) an expression can be obtained for the time derivative of the expectation value of an arbitrary operator $\hat{O}$ that acts on the system:
\begin{eqnarray}
\frac{\partial\left<\hat{O}\right>}{\partial t} = &&i\left<\left[\hat{H},\hat{O}\right]\right> \nonumber \\
&&+  \frac{1}{2}\sum_j\left(\left<\left[\hat{C}^\dagger_j,\hat{O}\right]\hat{C}_j\right>  + \left<\hat{C}^\dagger_j\left[\hat{O},\hat{C}_j\right]\right>\right).
\label{eq:TimeOp}
\end{eqnarray} 
Although this is a general picture, from now on we will focus on the class of systems consisting of  interacting bosons hopping on a lattice. Within the formalism of second quantization,  we introduce the bosonic annihilation operators  $\left\{\hat{a}_i\right\}$ and  creation operators $\left\{\hat{a}_i^\dagger\right\}$, where $i$ indicates the mode index. Such operators satisfy the bosonic commutation rules $
\left[\hat{a}_i,\hat{a}_j\right] = \left[\hat{a}_i^{\dagger},\hat{a}_j^\dagger\right] = 0; $ and $
\left[ \hat{a}_i,\hat{a}_j^\dagger \right] = \delta_{i,j}$
with $\delta_{i,j}$ the Kronecker delta. 

We now consider a Hamiltonian $\hat{H}$ that consists of a number conserving part $\hat{H}_{Sys}$ and a drive term $\hat{H}_{Drive}$. The particle conserving part typically contains one- and two-particle terms, namely: 
\begin{equation}
\hat{H}_{Sys} = \sum_{i,j}h^{(1)}_{i,j}\hat{a}_i^\dagger\hat{a}_j +  \sum_{i,j}h^{(2)}_{i,j}\hat{a}_i^\dagger\hat{a}_j^\dagger\hat{a}_i\hat{a}_j,
\label{eq:HamNumCon}
\end{equation} 
where the sums run over all the modes of the system. The matrix $ h^{(1)}_{i,j}$ can be used to describe the kinetic energy and external potentials while the matrix $h^{(2)}_{i,j}$ introduces interactions. 

For the drive term a coherent pump field is considered:
\begin{equation}
\hat{H}_{Drive} = \sum_{i}\left(F_i\hat{a}_i^\dagger+F_i^*\hat{a}_i\right),
\label{eq:HamDrive}
\end{equation} 
with $F_i$ the drive amplitude for mode $i$. Note that this is written in the frame rotating at the drive frequency (in the rotating rame, the system Hamiltonian  conserves the general form in Eq. (\ref{eq:HamNumCon})). 

For the environment we assume that the bosonic excitations can escape to a bath at temperature zero. This is described by jump operators that are proportional to the annihilation operators $\hat{C}_j=\sqrt{\gamma_j}\hat{a}_j$, with $\gamma_j$ the corresponding dissipation rate. It is straightforward to also introduce jump operators that populate the modes incoherently (proportional to the creation operators), which can describe a non-resonant pump or the thermal excitations of the environment. Also other types of dissipation processes can be considered, such as pure dephasing \cite{PhysRevA.84.032301}.

\subsection{Equations of motion for the correlation functions}
All the observables that are local in time can be written in terms of the correlation functions $\left<\prod_i\hat{a}_i^{\dagger n_i}\hat{a}_i^{m_i}\right>$. Applying expression (\ref{eq:TimeOp}),  we get  the equations of motion for the correlation functions. For the one-particle  term of the number conserving part (\ref{eq:HamNumCon}),  we find:
\begin{widetext}
\begin{eqnarray}
\left[\hat{H}^{(1)}_{Sys},\prod_k\hat{a}_k^{\dagger n_k}\hat{a}_k^{m_k}\right] = && \sum_{i,j(i\neq j)}h^{(1)}_{i,j}\left(\prod_{k\neq i,j}\hat{a}_k^{\dagger n_k}\hat{a}_k^{m_k}\right)\left(n_j\hat{a}_i^{\dagger n_i+1}\hat{a}_i^{m_i}\hat{a}_j^{\dagger n_j - 1}\hat{a}_j^{m_j} - m_i\hat{a}_i^{\dagger n_i}\hat{a}_i^{m_i-1}\hat{a}_j^{\dagger n_j}\hat{a}_j^{m_j+1}\right) \nonumber \\
&& + \sum_{i}h^{(1)}_{i,i}(n_i - m_i)\left(\prod_{k}\hat{a}_k^{\dagger n_k}\hat{a}_k^{m_k}\right).
\end{eqnarray} 
The two-particle part gives: 
\begin{eqnarray}
\left[\hat{H}^{(2)}_{Sys},\prod_k\hat{a}_k^{\dagger n_k}\hat{a}_k^{m_k}\right] 
= &&
\sum_{i,j}h^{(2)}_{i,j}\left(\prod_k\hat{a}_k^{\dagger n_k}\hat{a}_k^{m_k}\right) \left[ \left(n_i-m_i\right)\hat{a}_j^{\dagger}\hat{a}_j + \left(n_j-m_j\right)\hat{a}_i^{\dagger}\hat{a}_i \right.  \nonumber\\
&&\left. - m_j\left(m_i  - \delta_{i,j}\right) + n_i\left(n_j - \delta_{i,j}\right) \right].
\end{eqnarray} 
The part describing the coherent drive results in:
\begin{equation}
\left[\hat{H}_{Drive},\prod_k\hat{a}_k^{\dagger n_k}\hat{a}_k^{m_k}\right] 
= 
\sum_{i} \left( \prod_{k \neq i}\hat{a}_k^{\dagger n_k}\hat{a}_k^{m_k}\right) \left(n_iF_i^*\hat{a}_i^{\dagger n_i-1}\hat{a}_i^{m_i}-m_iF_i\hat{a}_i^{\dagger n_i}\hat{a}_i^{m_i - 1}\right).
\end{equation} 
Finally, for the term corresponding to the dissipation in expression (\ref{eq:TimeOp}), we need:
\begin{eqnarray}
\hat{a}_i^\dagger \left[\prod_k\hat{a}_k^{\dagger n_k}\hat{a}_k^{m_k},\hat{a}_i\right] &&= - n_i \prod_{k}\hat{a}_k^{\dagger n_k}\hat{a}_k^{m_k} \\
\left[\hat{a}_i^\dagger,\prod_k\hat{a}_k^{\dagger n_k}\hat{a}_k^{m_k}\right]\hat{a}_i &&= - m_i \prod_{k}\hat{a}_k^{\dagger n_k}\hat{a}_k^{m_k}.
\end{eqnarray} 
Summing all the terms, we find for the time derivative of an arbitrary correlation function:
\begin{eqnarray}
\frac{\partial\left<\prod_k\hat{a}_k^{\dagger n_k}\hat{a}_k^{m_k}\right>}{\partial t} =  
&& i\sum_{i,j}h^{(1)}_{i,j}\left(n_j\left. \left<\prod_k\hat{a}_k^{\dagger n_k}\hat{a}_k^{m_k}\right>\right|_{n_i\rightarrow n_i+1 \atop n_j\rightarrow n_j-1}  - m_i\left. \left<\prod_k\hat{a}_k^{\dagger n_k}\hat{a}_k^{m_k}\right>\right|_{m_j\rightarrow m_j+1 \atop m_i\rightarrow m_i-1} \right) \nonumber\\
&&+ i\sum_{i,j}h^{(2)}_{i,j}\left\{ \left(n_i-m_i\right) \left.\left<\prod_k\hat{a}_k^{\dagger n_k}\hat{a}_k^{m_k}\right>\right|_{n_j\rightarrow n_j+1 \atop m_j\rightarrow m_j+1}  + \left(n_j-m_j\right)\left.\left<\prod_k\hat{a}_k^{\dagger n_k}\hat{a}_k^{m_k}\right>\right|_{n_i\rightarrow n_i+1 \atop m_i\rightarrow m_i+1}\right.  \nonumber\\
&&\left. +\left[n_i\left(n_j - \delta_{i,j}\right) - m_j\left(m_i - \delta_{i,j}\right) \right]\left<\prod_k\hat{a}_k^{\dagger n_k}\hat{a}_k^{m_k}\right> \right\} \nonumber \\
&&+i\sum_{i} \left(n_iF_i^*\left.\left<\prod_k\hat{a}_k^{\dagger n_k}\hat{a}_k^{m_k}\right>\right|_{n_i\rightarrow n_i-1}-m_iF_i\left.\left<\prod_k\hat{a}_k^{\dagger n_k}\hat{a}_k^{m_k}\right>\right|_{m_i\rightarrow m_i-1}\right) \nonumber \\
&& - \frac{1}{2}\sum_i \gamma_i\left(n_i + m_i\right) \left<\prod_k\hat{a}_k^{\dagger n_k}\hat{a}_k^{m_k}\right>,
\label{Eq:CorrFu}
\end{eqnarray} 
where the following notation has been introduced:
\begin{eqnarray}
\left. \left<\prod_k\hat{a}_k^{\dagger n_k}\hat{a}_k^{m_k}\right>\right|_{n_i\rightarrow n} = \left<\hat{a}_i^{\dagger n}\hat{a}_i^{m_i}\prod_{k\neq i}\hat{a}_k^{\dagger n_k}\hat{a}_k^{m_k}\right>.
\end{eqnarray} 
\end{widetext}
To make the notation more compact and valid for all the integer indexes, note that formally there are a few terms with negative exponents, but they are of course equal to zero because the pre-factors multiplying them vanish as it can be readily verified. Eq. (\ref{Eq:CorrFu}) corresponds to a linear set of equations for the correlation functions. Due to the two-particle part $\hat{H}^{(2)}_{Sys}$ in the Hamiltonian each correlation function is a function of higher order correlation functions, thus resulting in a hierarchy with an infinite set of coupled equations. If one is interested in the properties of the steady-state the time derivatives can be set to zero. 

In order to find a numerical solution for the correlation functions, a truncation scheme  of the infinite hierarchy of equations must be introduced. We will introduce a global cutoff on the correlation order $N_C$, i.e., we will consider only the equations of motion for correlation functions with $\sum_i n_i \leq N_C$ and $\sum_i m_i \leq N_C$. Note that this is different from a local cutoff, as for example considered in Ref. \cite{0953-4075-46-22-224023}, for which all the correlation functions with $n_i\leq N_C$ and $m_i\leq N_C$ are retained, for all $i$. In the case of multiple modes the global cutoff allows to capture higher order correlation functions with a smaller set of equations.

It is clear that if we consider the correlation functions up to the order $N_C$, in their equations a dependence on higher order correlation functions remains. In the following, we propose two complementary schemes to deal with them.

\subsection{Global 'hard' cutoff}
{ The first rather straightforward possibility corresponds to setting these correlation functions equal to zero, which we denote as a hard cutoff.
Namely, we impose $\left <\prod_k\hat{a}_k^{\dagger n_k}\hat{a}_k^{m_k}\right> = 0$ for $\sum_i n_i > N_C$ or $\sum_i m_i >  N_C$.}
This cutoff is expected to be adequate in the limit of a small number of excitations. 

This scheme result in a total number of correlation functions $N^{(HC)}$ for which the equations of motion have to be solved, given by:
\begin{eqnarray}
N^{(HC)} = \left[\sum_{i=0}^{N_C} {N_S + i -1 \choose i}\right]^2,
\label{NumHC}
\end{eqnarray} 
with $N_S$ the total number of modes and ${N_S + i -1 \choose i}$ the binomial coefficient representing the number of possibilities to choose $i$ elements from a set of size $N_S$, allowing for repetition. This expression can be understood as follows: for $N_S$ modes there are ${N_S + i -1 \choose i}$ possibilities to choose a set $\left\{ n_k \right\}$ with $\sum_k n_k = i$, thus there are $\sum_{i=0}^{N_C} {N_S + i -1 \choose i}$ possible sets with $\sum_i n_i \leq N_C$. The number of correlation functions with $\sum_i n_i \leq N_C$ and $\sum_i m_i \leq N_C$ is the square of this number, leading to Eq. (\ref{NumHC}). For a large system size $N_S$ the dominant contribution to Eq. (\ref{NumHC}) is:
\begin{eqnarray}
N^{(HC)} &&\propto N_S^{2N_C} 
\label{NumHCDom}
\end{eqnarray} 
Hence, for a given cutoff order $N_C$ the number of equations grows according to a power law as a function of the size of the system. This is a consequence of the global cutoff: for a local cutoff, instead, the number of equations grows exponentially with the system size. This shows that truncation schemes based on a global cutoff can be very efficient for problems that involve a low order of correlation.

\subsection{Global self-consistent 'soft' cutoff}

When the steady-state is in a coherent state,  the correlation functions decouple: $\langle\prod_k\hat{a}_k^{\dagger n_k}\hat{a}_k^{m_k}\rangle = \prod_k\langle\hat{a}_k^{\dagger}\rangle^{n_k}\langle\hat{a}_k\rangle^{m_k}$. Hence,  in a regime where the number of excitations is large and the system is not so far from a coherent state, we are led to consider the following 'soft' cutoff condition:
 \begin{eqnarray}
\left<\prod_k(\hat{a}_k^{\dagger} - \langle \hat{a}_k \rangle^*)^{n_k}(\hat{a}_k - \langle \hat{a}_k \rangle)^{m_k}\right> = 0,
\label{softCut}
\end{eqnarray} 
for $\sum_i n_i > N_C$ or $\sum_i m_i > N_C$.  Note that this equation gives an expression for $ \left<\prod_k\hat{a}_k^{\dagger n_k}\hat{a}_k^{m_k}\right>$ with $\sum_i n_i> N_C$ or $\sum_i m_i > N_C$ in terms of lower order correlation functions and thus closes the system of equations. This scheme can be thought of as an expansion in correlations around a coherent state, thus allowing the study of correlations in regimes with a high photon density that are unattainable with numerical approaches based on a cutoff in number of photons. Note that if the cutoff is set to $N_c=1$ this reduces to the non-equilibrium Gross-Pitaevskii approximation \cite{RevModPhys.85.299}. A consequence of the soft cutoff is that the set of equations of motion becomes nonlinear and has to be solved self-consistently for $\left\{\langle \hat{a}_k \rangle\right\}$. We note that this approach with the soft cutoff is related to a cumulant expansion, as for example pursued in Ref. \cite{PhysRevA.41.3847}.

A consequence of the soft cutoff is that the set of equations of motion becomes nonlinear and has to be solved self-consistently for $\left\{\langle \hat{a}_k \rangle\right\}$. In practice this is obtained by using the Gross-Pitaevskii prediction (corresponding to $N_C = 1$) for the $\left\{\langle \hat{a}_k \rangle\right\}$ as an initial guess. These values are updated by solving the system of equations with the 'old' values for $\left\{\langle \hat{a}_k \rangle\right\}$ in (\ref{softCut}) which results in a new set $\left\{\langle \hat{a}_k \rangle\right\}$. This procedure is then iterated until convergence is reached. For the regimes considered in this paper this is obtained already after a few iterations. 

 In this case, the total number of correlation functions $N^{(SC)}$ that have to be solved self-consistently is:
\begin{eqnarray}
N^{(SC)} = \left[\sum_{i=0}^{N_C + 1} {N_S + i -1 \choose i}\right]^2.
\label{NumSC}
\end{eqnarray} 
This can be understood by following the same reasoning as presented in the previous subsection for a hard cutoff and adding the number of Eqs. (\ref{softCut}) that are needed to close the system of equations.  Similar to Eq. (\ref{NumHCDom}) for a hard cutoff, the dominant contribution to Eq. (\ref{NumSC}) for a large system size $N_S$ is:
\begin{eqnarray}
N^{(SC)} \propto N_S^{2N_C + 2} .
\end{eqnarray} 
Since this approach results in a larger set of equations with respect to the hard cutoff scheme it is limited to smaller systems and/or a smaller cutoff $N_C$. 

\section{Driven-dissipative Bose-Hubbard model \label{DDBH}}
As an example we apply the presented method to the driven-dissipative Bose-Hubbard model. The corresponding system Hamiltonian is (in the frame rotating at the pump frequency):
 \begin{eqnarray}
\hat{H}_{BH} = && \sum_{i}\left(-\Delta\hat{a}_i^\dagger\hat{a}_i  + \frac{U}{2}\hat{a}_i^\dagger\hat{a}_i^\dagger\hat{a}_i\hat{a}_i\right) \nonumber \\
&& -J\sum_{\left<i,j\right>}\left(\hat{a}_i^\dagger\hat{a}_j + h.c.\right).
\end{eqnarray} 
The parameters are the detuning $\Delta = \omega_p - \omega_c$ between the pump and the cavity frequency, the on-site interaction strength $U$ and the hopping parameter $J$. The sum $\sum_{\left<i,j\right>}$ is restricted to the nearest neighbours. Note that this Hamiltonian indeed has the form considered in Eq. (\ref{eq:HamNumCon}). Furthermore, in the rotating frame the coherent drive is described by Eq. (\ref{eq:HamDrive}). 

We will consider one-dimensional chains and two-dimensional arrays, both with periodic boundary conditions. The length of a chain is denoted as $L = Na$, with $N$ the number of sites and $a$ the lattice constant. The linear ($U = 0$) system Hamiltonian is diagonal in the reciprocal space, with $k_n = n 2\pi/L$ the allowed wave numbers with $ n \in \{0,1, ..., N-1\}$ and contains resonances at  $\Delta = -2J\cos(ak_n)$. For the two dimensional clusters, $N_x$ sites in the $x$-direction and $N_y$ sites in the $y$-direction are considered (denoted as a $N_x \times N_y$ cluster). The allowed $\vec{k}$-vectors are: $(k_{n_x},k_{n_y}) = 2\pi/a\times(n_x/N_x,n_y/N_y)$, with $ n_x \in \{0,1, ..., N_x-1\}$ and $ n_y \in \{0,1, ..., N_y-1\}$. In this case the linear resonances are at $\Delta = -2J\cos\left(ak_{n_x} \right) - 2J\cos\left(ak_{n_y} \right)$. For a homogeneous system with uniform drive only the modes corresponding to $n = 0$ (1D) or $(n_x,n_y) = (0,0)$ (2D) are driven, the other modes can only get populated through nonlinear interactions.

In the following, we will calculate the local density $n_i$ and the normalised second-order correlation function $g^{(2)}_{i,j}$, defined as:
 \begin{eqnarray}
n_i = \langle \hat{a}^{\dagger}_i\hat{a}_i \rangle; \\
g^{(2)}_{i,j} = \frac{\langle \hat{a}^{\dagger}_i\hat{a}^{\dagger}_j\hat{a}_j\hat{a}_i \rangle}{n_in_j}.
\end{eqnarray} 
Since we will consider translationally invariant systems the value of $i$ is not important and will be omitted for the density, i.e. $n_i= n$.

\section{Numerical results}
In this section we present results for the steady-state of the driven-dissipative Bose-Hubbard model with a homogeneous drive ($F_i = F$) in 1 and 2 dimensions with periodic boundary conditions. Note that in 1D the periodic boundary conditions can be realised experimentally by arranging the resonators in a ring geometry \cite{PhysRevLett.112.210405, PhysRevX.5.011034}. We will consider two complementary sets of system parameters that are ideally suited for the two presented truncation schemes. 

The convergence of the results as a function of $N_C $ has been verified numerically for the smaller system sizes. Indeed the considered regimes are well described by the presented truncation schemes already with a cutoff $N_C = 2$.  
For the largest considered system sizes, the accuracy of the results have been successfully benchmarked by comparison to calculations based on the corner-space renormalization method \cite{PhysRevLett.115.080604} (convergence of the results has been carefully checked by increasing the dimension $M$ of the corner space, with relative errors below $1 \%$. )

\begin{figure}[h]
  \includegraphics[scale=0.9]{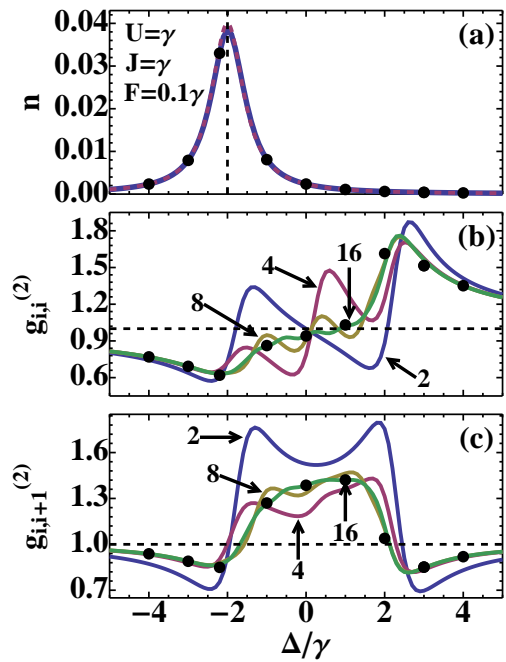}
  \caption{\label{Fig1} The local density (a), the on-site second-order correlation $g_{i,i}^{(2)}$ (b) and the nearest neighbour $g_{i,i+1}^{(2)}$ as a function of the detuning $\Delta$ (in units of $\gamma$) for the driven dissipative Bose-Hubbard 1D chain with periodic boundary conditions. For the density there is no significant difference for different lattice sizes and only a very small difference with the linear spectrum (shown as a dashed curve but very hard to distinguish). For the correlation functions the length of the chain is denoted in the figures. The system parameters are: $J = \gamma$, $U=1\gamma$, $F = 0.1\gamma$. The black dots are results obtained with the corner-space renormalization method\cite{PhysRevLett.115.080604}  for the largest considered system of 16 coupled cavities (with a corner space dimension $M = 1000$; relative errors below $1\%$).}
\end{figure}

\begin{figure}[h]
  \includegraphics[scale=0.85]{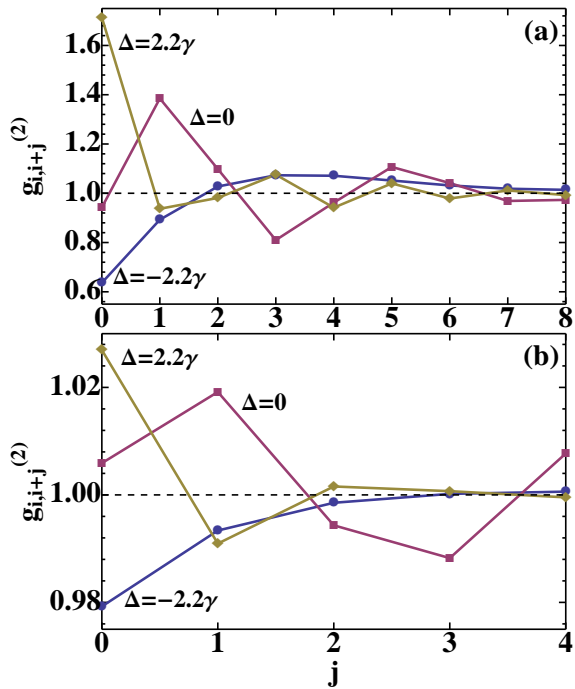}
  \caption{\label{Fig2} The normalised second-order correlation function $g_{i,i+j}^{(2)}$ as a function of the distance $j$ for three values of the detuning ($\Delta/\gamma = -2.2$, $0$ and $2.2$). The system parameters are: $J = \gamma$, $U=\gamma$ and $F = 0.1\gamma$ in (a) and $J = \gamma$, $U=0.05\gamma$ and $F = 1.2\gamma$ in (b).}
\end{figure}

\subsection{One Dimensional Chains}
\subsubsection{Hard Cutoff \label{1DHC}}

\begin{figure}[h]
  \includegraphics[scale=0.9]{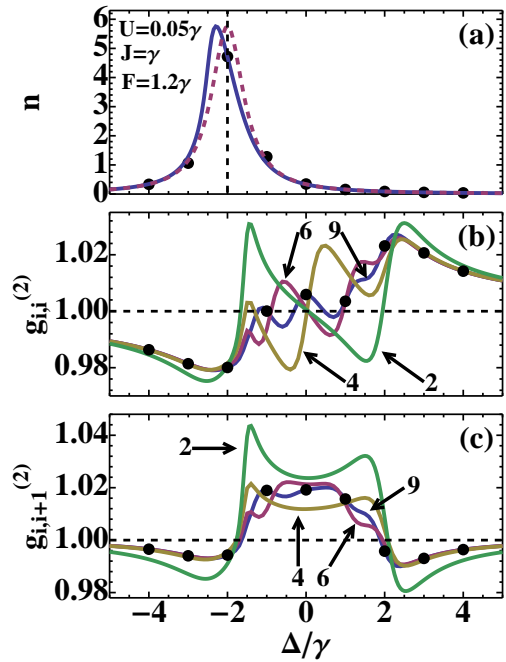}
  \caption{\label{Fig3} The local density (a), the on-site $g_{i,i}^{(2)}$ (b) and the nearest neighbour $g_{i,i+1}^{(2)}$ (c) second order correlation functions as a function of the detuning $\Delta$ for the driven dissipative Bose-Hubbard lattice in 1D with periodic boundary conditions. For the density there is practically no difference for different lattice sizes but the nonlinear shift with respect to the linear spectrum (dashed curve) can be clearly seen. For the correlation functions the size of the lattice is denoted in the figures. The system parameters are: $J = \gamma$, $U=0.05\gamma$, $F = 1.2\gamma$. The black dots are results obtained with the corner-space renormalization method  for the largest considered system of 9 coupled cavities (with a corner-space dimension $M$ = 500; relative error below $1\%$).}
\end{figure}

In Fig. \ref{Fig1} the results are presented for 1D chains with up to $16$ sites as a function of the detuning $\Delta$, for the system parameters $J = \gamma$, $U=\gamma$ and $F = 0.1\gamma$. Due to the small density, the hard cutoff truncation scheme has been considered. For the local density in Fig. \ref{Fig1} (a) only the result for a single cavity is presented since it is practically independent of the chain length and furthermore deviates only slightly from the linear spectrum (also presented in Fig. \ref{Fig1} (a)). 

The on-site normalized second order correlation function $g_{i,i}^{(2)}$ is presented in Fig. \ref{Fig1} (b) .  In contrast to the behavior of the density in Fig. \ref{Fig1} (a), the second-order correlation function dramatically depends on the chain length. For sufficiently negative detuning anti-bunching is observed and for sufficiently positive detuning we find bunching. This behavior is already well-known for a single cavity with a anti-bunching to bunching transition around the resonance \cite{PhysRevB.73.193306}. As the length is increased an overall attenuation of the bunching and anti-bunching is found. Moreover, an oscillating behavior appears at the detunings corresponding to the linear resonances at $\Delta = 2J\cos(ak_n)$. Note that since the coherent drive is homogeneous only the $n = 0$ mode is pumped and all other modes are only populated through scattering and are completely absent for the linear system. The amplitude of these oscillations decreases as the number of modes increases, indicating a bunching to anti-bunching transition around $\Delta = 0$ in the thermodynamic limit $N \rightarrow \infty$. 
Note that the results are in excellent agreement with the predictions of the corner-space renormalization method (black dots).

In Fig. \ref{Fig1} (c) the nearest neighbour normalised second order correlation function $g_{i,i+1}^{(2)}$ is presented as a function of the detuning. This reveals that $g_{i,i+1}^{(2)} < 1$ for large detuning $|\Delta| > 2J$, with a minimum just outside the edges of the linear energy band at $\Delta = \pm2J$. For small detunings we find $g_{i,i+1}^{(2)} > 1$. Similar to the results for the local $g_{i,i}^{(2)}$ in Fig. \ref{Fig1} (b), the overall values of $g_{i,i+1}^{(2)}$ decrease as the chain length is increased. Also here, an oscillating behavior is observed for the finite systems at the detunings corresponding to the linear eigenmodes, with a decreasing amplitude as the length is increased.

In Fig. \ref{Fig2} (a) the normalised second order correlation function $g_{i,i+j}^{(2)}$ is presented as a function of the distance $j$ for three values of the detuning ($\Delta/\gamma = -2.2$, $0$ and $2.2$). The results show an overall decay to $g_{i,i+j}^{(2)} \rightarrow 1$ for $j \rightarrow \infty$, with superimposed oscillations. Similar oscillations in the non-local $g_{i,i+j}^{(2)}$ have been predicted for various one dimensional driven-dissipative nonlinear photonic system \cite{PhysRevLett.104.113601, PhysRevLett.115.143601,PhysRevLett.110.163605} and they typically indicate a regular ordering of the photons.   

\subsubsection{Soft Cutoff \label{1DSC}}
We now perform a similar analysis as presented in the previous subsection, but for the system parameters $J = \gamma$, $U=0.05\gamma$, $F = 1.2\gamma$. This corresponds to a relatively large photon density and is ideally suited for the soft cutoff truncation scheme. As discussed before, due to the soft cutoff we are limited to smaller chain sizes as compared with the hard cutoff and we consider chains with up to $9$ sites. In Fig. \ref{Fig3} (a) the density is presented as a function of the detuning for which the dependence on the chain size is again negligible. Note that the resulting density is much larger than the one found in Fig.  \ref{Fig1} (a) and the nonlinear shift of the resonance peak with respect to the linear system is now clearly visible. In Fig. \ref{Fig3} (b) and (c) the local and nearest neighbour normalised second order correlation functions are presented. Due to the larger photon density the system is closer to a coherent state and the normalised correlation functions are much closer to $1$ with respect to the results in Fig. \ref{Fig1}. However, the same qualitative behavior is found as in Fig. \ref{Fig1}. In Fig.  \ref{Fig2} (b) the dependence of the $g_{i,i+j}^{(2)}$ on the distance $j$ is presented which is also qualitatively similar to the result in Fig. \ref{Fig2} (a). Note that also here an excellent agreement is found with the results obtained with the corner-space renormalization method for the largest system size (black dots).

\subsection{Two Dimensional Clusters}

\begin{figure}[h]
  \includegraphics[scale=0.9]{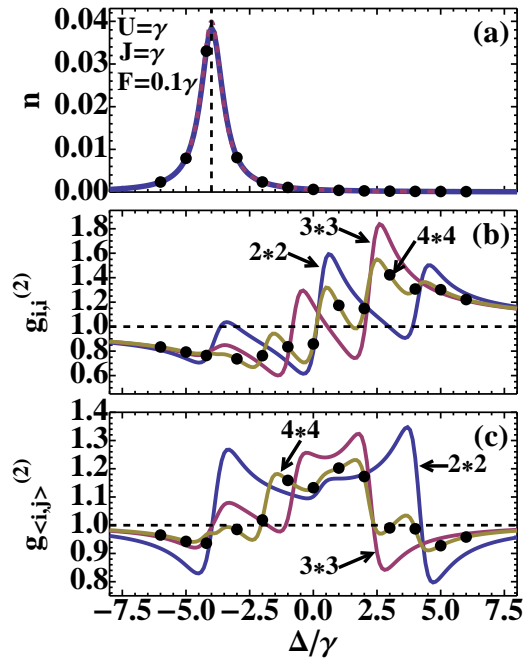}
  \caption{\label{Fig4} The density (a), the on-site $g_{i,i}^{(2)}$ (b) and the nearest neighbour $g_{<i,j>}^{(2)}$ (c) second order correlation functions as a function of the detuning $\Delta$ for the driven dissipative Bose-Hubbard lattice in 2D with periodic boundary conditions. The geometry of the considered clusters are characterized as $N_x*N_y$. For the density there is practically no difference for different lattice sizes and only a small difference with the linear spectrum (shown as a dashed curve, practically indistinguishable). The system parameters are: $J = \gamma$, $U=1\gamma$, $F = 0.1\gamma$. The black dots are results obtained with the corner-space renormalization method for the largest considered system of 16 coupled cavities (with a corner space dimension $M = 1500$; relative error below $1\%$) .}
\end{figure}

\subsubsection{Hard Cutoff}
Here, we again consider the system parameters $J = \gamma$, $U=0.1\gamma$, $F = \gamma$ and the hard cutoff truncation scheme, as in Sec. \ref{1DHC}, but now for 2D clusters with up to 16 sites arranged in a square geometry ($N_x = N_y$). In Fig. \ref{Fig4} (a) the density is presented as a function of the detuning $\Delta$ which is again practically independent of the size of the cluster and indistinguishable from the linear result. In Fig. \ref{Fig4} (b) and (c) the local and nearest neighbour second order correlation functions are presented as a function of the detuning $\Delta$. This again exhibits the same qualitative behavior as found for 1D chains with an anti-bunching to bunching transition that becomes less pronounced as the cluster size is increased and superimposed oscillations at the linear resonances. These oscillations again diminish for increasing size indicating a anti-bunching to bunching transition around $\Delta = 0$ in the thermodynamic limit $N_x, N_y \rightarrow \infty $. For the non-local $g_{\langle i,j\rangle}^{(2)}$ ($\langle i,j\rangle$ denotes nearest neighbours) we again find $g_{\langle i,j\rangle}^{(2)}<1$ far detuned from the linear resonances ($|\Delta| > 4J$), with a  minimum just outside the edge of the linear energy band $\Delta = \pm 4J$; instead,   $g_{\langle i,j\rangle}^{(2)} > 1$ for detuning within the linear energy band.

\subsubsection{Soft Cutoff}
As before in Sec. \ref{1DSC}, the system parameters $J = \gamma$, $U=0.05\gamma$, $F = 1.2\gamma$ are considered and the system is studied with the soft cutoff truncation scheme for clusters with up to 9 sites arranged in a $3 \times 3$ geometry. In Fig. \ref{Fig4} (a) the density is presented which is again practically independent of the size of the cluster and much larger with respect to the previous subsection with a clear nonlinear shift of the resonance peak. Also the local and nearest neighbour second order correlation functions in Fig. \ref{Fig4} (b) and (c) exhibit the same qualitative behavior as discussed before.
Note that the results presented in both Figs. \ref{Fig3} and \ref{Fig4} for the largest considered system sizes are in excellent agreement with the predictions of the corner-space renormalization method.
Notice the reported results correspond to a mean-number of photons per site exceeding 5. Such larger number of bosons per site cannot be described by brute-force integration of the master equation in the Fock basis, because the Hilbert size dimension becomes prohibitively large.

\begin{figure}[h]
  \includegraphics[scale=0.9]{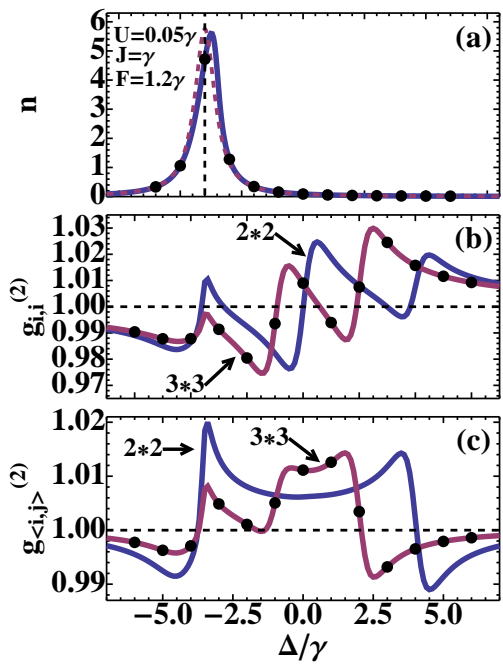}
  \caption{\label{Fig5} The density (a), the on-site $g_{i,i}^{(2)}$ (b) and the nearest neighbour $g_{<i,j>}^{(2)}$ (c) second order correlation functions as a function of the detuning $\Delta$ for the driven dissipative Bose-Hubbard lattice in 2D with periodic boundary conditions. The geometry of the considered clusters are characterized as $N_x*N_y$. For the density there is practically no difference for different lattice sizes and a clear nonlinear shift with respect to the linear spectrum (shown as a dashed curve). The system parameters are: $J = \gamma$, $U=0.05\gamma$, $F = 1.2\gamma$. The black dots are results obtained with the corner-space renormalization method  for the largest considered system of 9 coupled cavities (with a corner-space dimension $M = 800$; relative error below $1 \%$).}
\end{figure}

\section{Conclusions and outlook}
We have presented a numerical approach suitable for the description of few-body correlations in systems out of equilibrium, based on the equations of motion for the correlation functions. Two possible complementary truncation schemes based on a global cutoff were discussed, one ideally suited for small densities and the other for large densities. As an example we have applied this scheme to the driven-dissipative Bose-Hubbard model in both one- and two-dimensions with periodic boundary conditions and examined the fate of the bunching to anti-bunching transition in extended lattices.  An oscillating behavior is observed for the local second-order correlation function $g_{i,i}^{(2)}$ as a function of the detuning for intermediate system sizes. The oscillations tend to disappear as the size of the system increases. An extrapolation of this behavior leads to an overall bunching to anti-bunching transition around $\Delta = 0$ in the thermodynamic limit. For the nearest neighbor correlations we found $g_{<i,j>}^{(2)} < 1$ for detuning outside of the single-particle energy band and $g_{<i,j>}^{(2)} >1$ for detunings within the band. For 1D chains also the dependence of the correlations on the distance was studied, revealing an oscillating behavior typically observed for these systems.
The presented results were successfully benchmarked with the corner-space renormalization method \cite{PhysRevLett.115.080604} by comparing a representative set of points.

The presented results reveal that there are physical problems for which the considered truncation schemes can be more efficient with respect to other numerical approaches. An important advantage with respect to other truncation schemes is that for a given correlation order the number of equations increases as a power law versus the number of sites in contrast to the typical exponential dependence for truncation approaches with a local cutoff. Furthermore, the numerical implementation is rather simple and straightforward with respect to more advanced numerical schemes such as the ones based on matrix product operators (MPO) in 1D and the corner-space renormalization method (1D and 2D). The results presented in this paper for the largest considered systems are obtained in a computational time of the order of minutes on a desktop computer. Although with MPO approaches and  the  corner-space renormalization method it  is possible to tackle larger lattices and larger correlations, the present truncation schemes with global cutoff can be much faster (a couple of orders of magnitude for the lattices of moderate size considered in this paper).

The presented method can be generalized to various other models and physical situations. One can consider multiple modes per site, allowing for example the study of the driven-dissipative Jaynes Cummings model \cite{Houck:2012aa, PhysRevX.5.031028}, an opto-mechanical coupling \cite{PhysRevLett.112.013601, RevModPhys.86.1391} or the photon polarization \cite{PhysRevX.5.011034}. Also dissipative spin systems such as for example the dissipative XYZ model \cite{PhysRevLett.110.257204,2016arXiv160206553J} or collective phases with hard-core bosons\cite{2016arXiv160106857W} are a possible extension. Other possibilities are the consideration of spatially inhomogeneous systems with site dependent parameters (for example to study transport properties \cite{PhysRevA.91.053815}), other lattice geometries (for example to study the effect of geometric frustration \cite{PhysRevLett.115.143601, PhysRevA.93.043833}), other boundary conditions (for example closed or twisted boundary conditions \cite{PhysRevLett.112.210405} or a (cluster) Gutzwiller mean field approach \cite{PhysRevLett.110.233601, PhysRevA.90.063821, 2016arXiv160206553J}) or an incoherent population of the modes (for example thermal excitations or an incoherent drive). Also regimes with higher order correlations can be examined which however limits the size of the considered lattices. If one is interested in the time-dependent correlation functions the coupled differential equations of motion (\ref{Eq:CorrFu}) can be solved numerically with the presented truncation schemes. This allows to examine dynamical phenomena such as the dynamical hysteresis typically encountered in the optical bistability regime \cite{PhysRevA.93.033824}.   

We thank N. Bartolo, T. Lacroix, F. Minganti, R. Rota for a critical reading.
We  acknowledge  support  from  ERC  (via  the Consolidator  Grant ``CORPHO'' No.  616233),  from ANR (Project QUANDYDE No. ANR-11-BS10-0001), and from Institut Universitaire de France. A. L. B.  acknowledges support from the SFB TRR/21 and the EU Integrating project SIQS

\bibliography{manusc}

\end{document}